\documentclass[12pt]{iopart}

\expandafter\let\csname equation*\endcsname\relax
\expandafter\let\csname endequation*\endcsname\relax
\usepackage{adjustbox}
\usepackage{graphicx}
\usepackage[caption=false, font=normalsize, labelfont=sf, textfont=sf]{subfig}

\usepackage{mathtools}

\begin{document}

\title{Ultra-efficient DC-gated all-optical graphene switch}

\author{Mohammed Alaloul*, Khalil As'ham, Haroldo T. Hattori, and Andrey E. Miroshnichenko*}

\address{School of Engineering and Information Technology, University of New South Wales, Canberra, ACT 2600, Australia}
\ead{m.alaloul@unsw.edu.au and andrey.miroshnichenko@unsw.edu.au}
\vspace{10pt}
\begin{indented}
\item[]August 2017
\end{indented}

\begin{abstract}
The ultrafast response and broadband absorption of all-optical graphene switches are highly desirable features for on-chip photonic switching. However, because graphene is an atomically thin material, its absorption of guided optical modes is relatively low, resulting in high saturation thresholds and switching energies for these devices. To boost the absorption of graphene, we present a practical design of an electrically-biased all-optical graphene switch that is integrated into silicon slot waveguides, which strongly confine the optical mode in the slotted region and enhance its interaction with graphene. Moreover, the design incorporates a silicon slab layer and a hafnia dielectric layer to electrically tune the saturation threshold and the switching energy of the device by applying DC voltages of \textless 0.5$\,$V. Using this device, a high extinction ratio (ER) of 10.3$\,$dB, a low insertion loss (IL) of $<$0.7$\,$dB, and an ultra-efficient switching energy of 79$\,$fJ/bit at 0.23$\,$V bias are attainable for a 40$\,$\textmu m long switch. The reported performance metrics for this device are highly promising and are expected to serve the needs of next-generation photonic computing systems.
\end{abstract}

%
\vspace{2pc}

\section{Introduction}

All-optical switches are devices that control the propagation of light in a photonic data link by applying a time-varying optical signal. These devices exhibit ultrafast switching times ($<$$1\,$ps) \cite{chai2017ultrafast}, and are thus utilized for optical information processing applications in data centers, high-performance computers, and optical networks. Nevertheless, there exists an energy-speed tradeoff, where switches with a $<$$1\,$ps switching time operate at switching energies in the range of several picojoules, and sub-pJ switches operate with \textgreater 1$\,$ps switching times \cite{ono2020ultrafast, alaloul2021low}. Recently, however, an energy-efficient (35$\,$fJ) and ultrafast (260$\,$fs) all-optical plasmonic graphene switch was demonstrated \cite{ono2020ultrafast}. The use of plasmonic nanostructures improves the efficiency of these devices, but they also introduce an excessive insertion loss. Next-generation telecom and datacom networks require low-energy ($<$$1 \,$pJ/bit) and low-insertion-loss ($<$$5 \,$dB) photonic devices \cite{giambra2021wafer}, which necessitates the quest for designing devices with alternative structures that satisfy these requirements. It was recently reported that the switching threshold of an all-optical graphene switch can be significantly reduced by applying a few volts bias \cite{opex}. Using this method, the chemical potential of graphene is electrostatically increased by the applied DC voltage, which enables a low-energy optical pump signal to fill the remaining conduction band states and saturate the absorption of graphene, hence achieving efficient all-optical switching. Electrical control of the saturable absorption threshold in graphene was experimentally reported in \cite{alexander2015electrically}. In \cite{opex}, the studied devices consisted of silicon nitride (Si$_3$N$_4$) wire waveguides, which have a relatively large cross-section and low confinement of guided light. Thus, instead of Si$_3$N$_4$ waveguides, we herein propose a novel design of an electrically-biased all-optical graphene switch that is integrated into silicon slot waveguides. Si slot waveguides confine and guide the optical mode in a relatively small cross-section, which strongly enhances its interaction with graphene. While several designs of all-optical graphene switches have been proposed in the literature \cite{ono2020ultrafast, alaloul2021low, opex, wang2020cmos}, these designs either suffer from high switching thresholds or a high insertion loss, and therefore do not meet the stringent demands of next-generation telecom and datacom networks. To overcome this challenge, we first boost the optical absorption of the graphene switch by integrating it into Si slot waveguides, where the guided optical mode is strongly enhanced and confined. Secondly, electrostatic gating is introduced into this configuration to tune the chemical potential, and thereby reduce the effective saturable absorption threshold of graphene. Because of the excellent absorption enhancement and the minimised saturable absorption threshold in this configuration, an ultra-low switching energy of 79$\,$fJ at an applied DC voltage of 0.23$\,$V is achieved for a 40$\,$\textmu m long all-optical switch, with a high extinction ratio ($ER$) of 10.3$\,$dB and a low insertion loss ($IL$) of $<0.7\,$dB. These features, combined with the ultrafast response of graphene within a timescale of $<150\,$fs \cite{alaloul2021plasmon}, are highly desirable for high-speed all-optical signal processing.

In the next section, the design methodology of the device and its structure are presented, and the modeling parameters for reproducing the results are provided. Following that, the device performance metrics are assessed by investigating its switching threshold, extinction ratio, insertion loss, and switching time. Then, these metrics are discussed and compared with those of other graphene-based devices that have been recently reported in the literature. Finally, the main findings of this report are summarized in a brief conclusion.

\section{Methods}

\subsection{Design}

\begin{figure} 
    \centering
  \subfloat[\label{1a}]{%
       \includegraphics[width=0.8\linewidth]{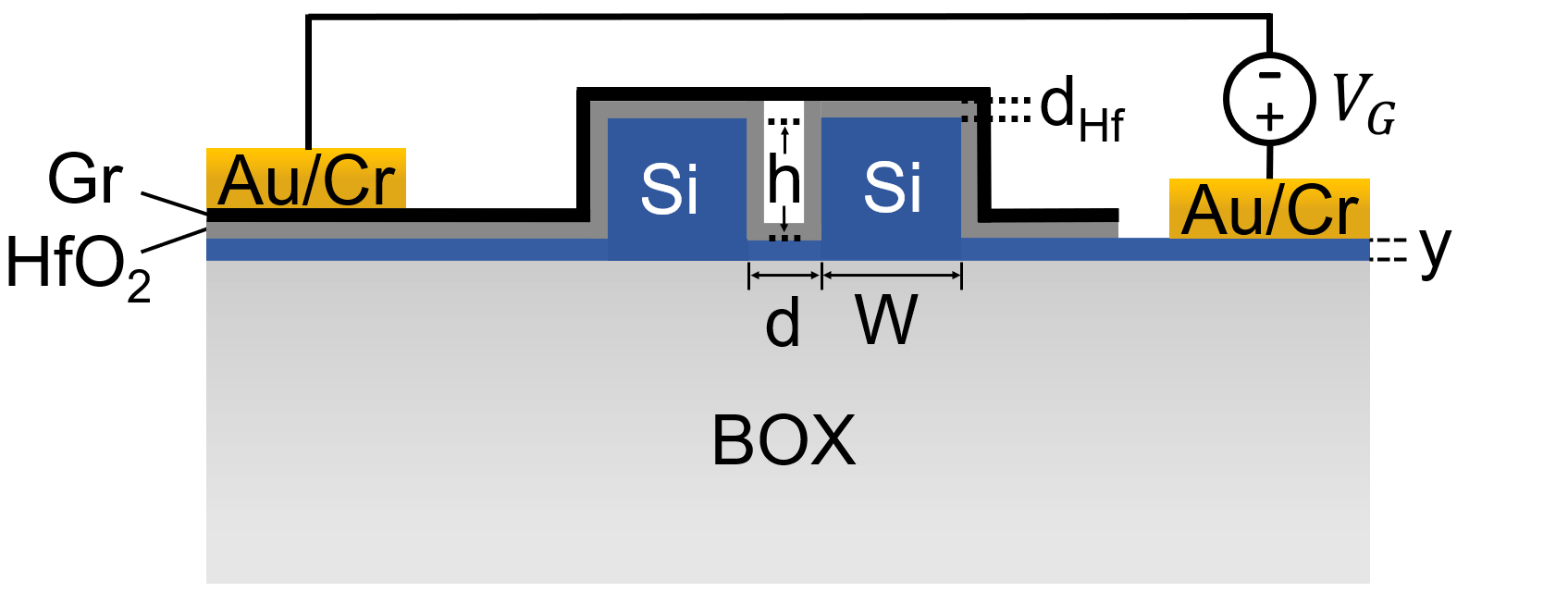}}
    \hfill
  \subfloat[\label{1b}]{%
        \includegraphics[width=1\linewidth]{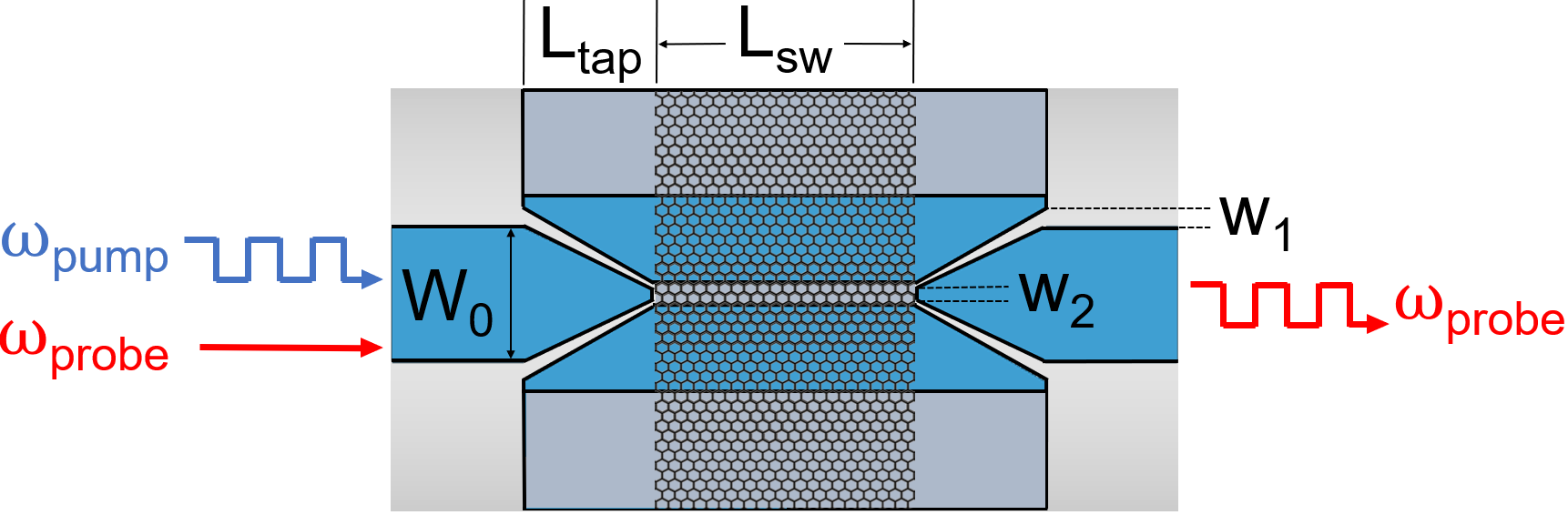}}
  \caption{(a) Front view and (b) top view of the on-chip switch. The pump signal modulates the probe signal. Gr: graphene, Si: silicon, HfO$_2$: hafnium dioxide (hafnia), BOX: buried oxide, $h$: rail thickness, $d$: slot width, $W$: rail width, $y$: slab thickness, $V_G$: DC gate voltage, $W_0$: strip waveguide width, $w_1$: separation distance, $L_{\text{tap}}$: taper length, $L_{\text{sw}}$: switch length, $w_2$: tip width.}
  \label{fig1} 
\end{figure}

\begin{figure} 
    \centering
        \subfloat[\label{slab_sweep}]{%
       \includegraphics[width=0.5\linewidth]{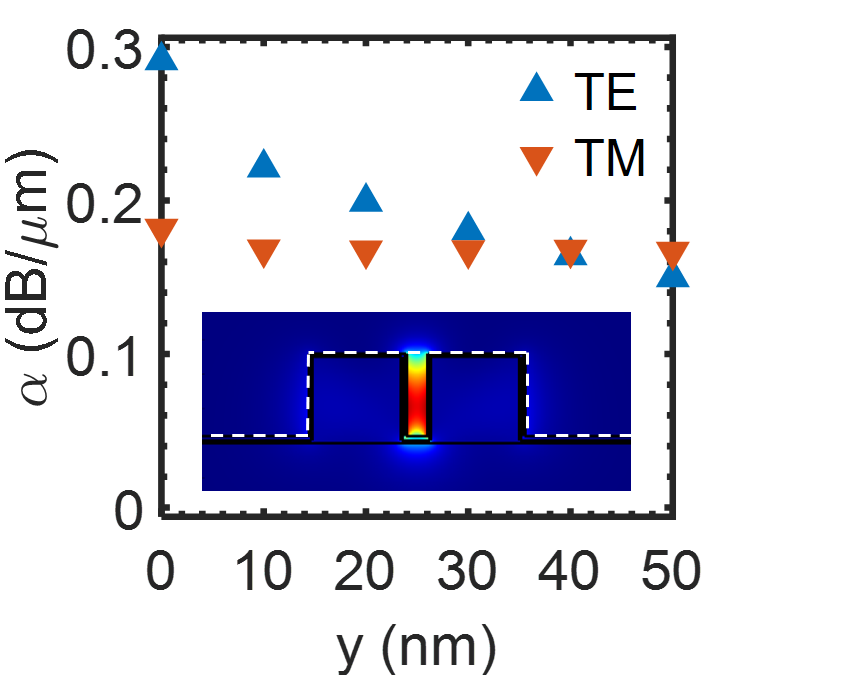}}
       \hfill
       \subfloat[\label{wavelength_sweep}]{%
        \includegraphics[width=0.5\linewidth]{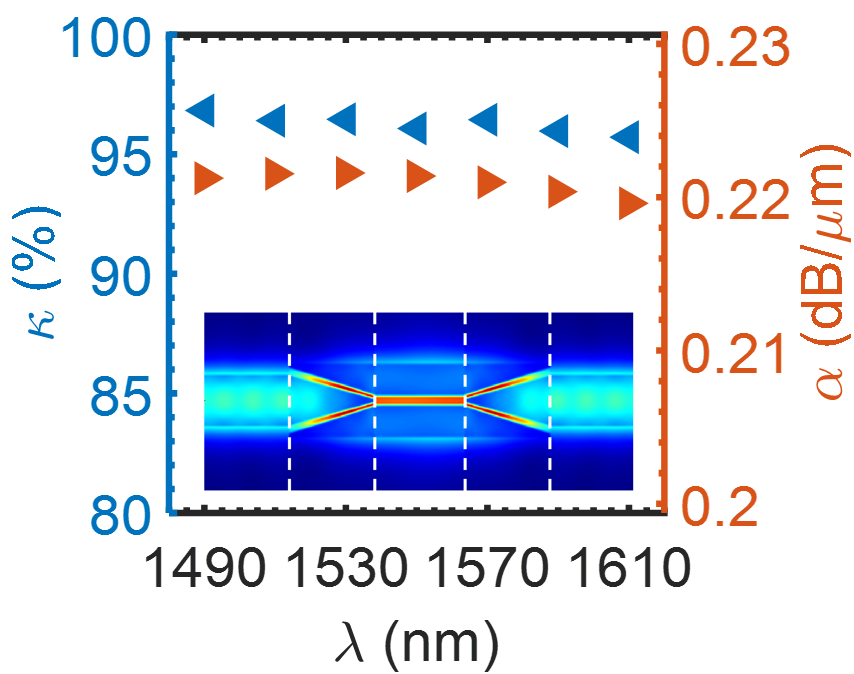}}
  \caption{(a) Propagation loss ($\alpha$) of the switch waveguide as a function of the slab height ($y$) for the transverse-electric (TE) and transverse-magnetic (TM) modes. The inset shows the propagating TE mode at $\lambda = 1550\,$nm and $y = 10\,$nm. (b) Coupling efficiency ($\kappa$) and propagation loss ($\alpha$) as a function of wavelength ($\lambda$). The inset shows the eigenmode expansion (EME) simulation of the TE mode at $\lambda = 1550 \,$nm.}
  \label{fig22} 
\end{figure}

The structure of the on-chip switch is illustrated in figure \ref{fig1}. It consists of a Si slot waveguide that is placed on top of a 10$\,$nm thick Si slab layer, which in turn sits atop a 2$\,$\textmu m thick buried oxide (BOX) layer. The Si rails are 240 $\,$nm thick, resulting in a 250$\,$nm total thickness of the Si layers. Silicon-on-insulator (SOI) substrates with 250$\,$nm thickness are commercially available. The device dimensions are optimized to achieve a high extinction ratio and energy-efficient switching at a compact footprint. The simulated ranges and optimal dimensions of the device are given in Table 1. Hafnia is an electrical insulator with a high dielectric constant ($\epsilon_{\text{Hf}} = 25$) \cite{robertson2004high}, making it a useful material for building field-effect transistors \cite{mikolajick2014doped}, memristors \cite{chen2021hafnium}, and optical modulators \cite{bonando2018all}. Thus, a 10$\,$nm thick hafnia layer is employed to form a capacitor configuration between graphene and Si. The waveguide supports a transverse-electric (TE) mode that has a computed propagation loss of $\alpha \approx 0.22 \,$dB/\textmu m at $\lambda = 1550\,$nm and $y=10\,$nm (see figure \ref{slab_sweep}). The propagation loss of the TE-mode is maximized when there is no Si slab, and it reduces when the slab thickness ($d$) increases. In this design, the Si slab is essential to bias graphene with $V_G$, so a minimum thickness of $10\,$nm is chosen because it yields the highest absorption for the TE-mode. 80$\,$nm Si slot waveguides can be realized in practice \cite{serna2014potential}, and can achieve a high graphene absorption in our configuration (see Appendix A). The computed TE-mode is shown in the inset of figure \ref{slab_sweep}. The transverse-magnetic (TM) mode has a lower propagation loss of $\alpha \approx 0.17 \,$dB/\textmu m at $\lambda = 1550\,$nm and $y=10\,$nm. The higher propagation loss of the TE-mode is explained by its stronger enhancement of the electric field profile \cite{leuthold2010nonlinear}, which boosts the effective absorption of graphene. As explained later, the switching mechanism of the device is based on the saturable absorption of graphene. Thus, a higher absorption is desirable to obtain higher switching efficiencies, which makes the TE-mode the favorable choice. The computed $\alpha$ values for the TE-mode at other wavelengths are presented in figure \ref{wavelength_sweep}. The waveguide modes are computed in Lumerical MODE, where these $\alpha$ values are obtained for graphene and hafnia layers with a total width of $W_{G} = 1.64\,$\textmu m, which is the effective absorbing width of graphene (see Appendix B). 

\begin{table}
\label{tab1}
\tabcolsep3.5pt
\caption{\bf Design parameters of the on-chip all-optical switch.}
\begin{tabular}{@{}*{11}{l}}
\br                              
\textbf{Parameter} & $h$ & $d_{\text{Hf}}$ & $y$ & $d$ & $W$ & $W_0$ & $w_1$ & $w_2$ & $L_{\text{sw}}$ & $L_{\text{tap}}$ \cr
\mr                              
\textbf{Range (\textmu m)} & $0.2-0.3$ & $0.01-0.1$ & $0-0.05$ & $0.04-0.12$ & $0.2-0.3$ & $-$ & $-$ & $-$ & $0-40$ & $0-10$ \cr
\mr
\textbf{Optimum (\textmu m)} & 0.24 & 0.01 & 0.01 & 0.08 & 0.25 & 0.4 & 0.06 & 0.03 & 40 & 8 \cr

\br
\end{tabular}
\end{table}

60$\,$nm thick and 300$\,$nm wide Au/Cr contacts supply the direct-current (DC) gate voltage ($V_G$) for tuning the carrier density of graphene, and the all-optical saturation threshold of graphene is electrically tuned as a result. The applied $V_G$ merely tunes the switching energy of the device, without switching the state of the propagating modes. Hence, the switching functionality of this device is all-optical, as explained in the next section. Moreover, the capacitor configuration that is presented in Fig. \ref{1a}, functions as an open circuit because the applied voltage is DC. As a result, the current ($I$) is zero, and thus the consumed electrical power ($P_E$) is zero based on $P_{E} = IV_G$ \cite{opex}. As explained in Appendix C, the left and right Au/Cr contacts are placed 0.8$\,$\textmu m away from the left and right Si rails, respectively, to ensure that they do not induce an ohmic loss. The computed propagation loss with the Au/Cr contacts at this separation distance is exactly the same as it is without their presence, which confirms that the Au/Cr films do not contribute to the propagation loss. To obtain a high switching efficiency at a compact footprint, we set the switch length ($L_{\text{sw}}$) to 40$\,$\textmu m. The slot waveguide is tapered to efficiently couple its guided mode with 250$\,$nm thick silicon wire waveguides, as illustrated in figure \ref{1b}. The taper design methodology is reported in \cite{wang2009ultracompact}. The tapered structure is simulated using the EigenMode Expansion (EME) Solver in Lumerical MODE, as shown in the inset of figure \ref{wavelength_sweep}. Simulations were conducted using the 3D solver with metal boundary conditions. The computed coupling efficiency and propagation loss as a function of wavelength are shown in figure \ref{wavelength_sweep}. High coupling efficiencies are obtained by tapering the waveguide structure, where the coupling efficiency ($\kappa$) at $\lambda = 1550\,$nm is as high as $\sim 96\%$. This result is in agreement with the one reported in \cite{wang2009ultracompact}, where a $\sim97$\% coupling efficiency has been demonstrated for an 8$\,\text{\textmu}$m long taper. The $\sim1$\% difference in our result is due to the presence of the hafnia and graphene layers, which introduce a slight mode mismatch, as confirmed by the power overlap analysis in Lumerical MODE.

\subsection{Modeling parameters}

The refractive index data of hBN are available in \cite{rah2019optical}. In Lumerical, graphene is simulated using the 2D model with a surface optical conductivity ($\Tilde{\sigma}$) \cite{alaloul2021plasmonic, hanson2008dyadic}:

\numparts
\begin{equation} \label{eq:1st}
    \Tilde{\sigma}(\omega, \Gamma, \text{\textmu}, T) = \Tilde{\sigma}_{intra}(\omega, \Gamma, \text{\textmu}, T) + \Tilde{\sigma}_{inter}(\omega, \Gamma, \text{\textmu}, T)
\end{equation}

\begin{equation} \label{eq:2nd}
     \Tilde{\sigma}_{intra}(\omega, \Gamma, \text{\textmu}, T) = \dfrac{-je^2}{\pi\hbar^2(\omega+j2\Gamma)}\int_{0}^{\infty}E\,(\dfrac{\partial f(E)}{\partial E} - \dfrac{\partial f(-E)}{\partial E}) \: dE
\end{equation}

\begin{equation} \label{eq:3rd}
     \Tilde{\sigma}_{inter}(\omega, \Gamma, \text{\textmu}, T) = \dfrac{je^2(\omega+j2\Gamma)}{\pi\hbar^2}\int_{0}^{\infty}\dfrac{f(-E) - f(E)}{(\omega+j2\Gamma)^2 - 4(E/\hbar)^2} \: dE
\end{equation}

\begin{equation} \label{eq:4th}
     f(E) = (e^{(E-\mu)/k_{B}T}+1)^{-1}
\end{equation}

\endnumparts

\noindent where $\Tilde{\sigma}_{intra}$ and $\Tilde{\sigma}_{inter}$ account for the surface optical conductivity due to intraband and interband absorption, respectively. $\Gamma$ is the scattering rate of graphene, $\omega$ is the angular frequency of incident photons, $e$ is the electron charge, $T$ is the operation temperature, $\hbar$ is the reduced Planck constant, $k_{B}$ is the Boltzmann constant, and $f(E)$ is the Fermi-Dirac distribution. The refractive index data of hafnia are taken from \cite{al2004optical}. In the optical simulation, the Au/Cr contact is simulated as a pure Au contact because the plasmonic response of Au is strong in the near-infrared (near-IR) wavelength band. The refractive index of Au are taken from \cite{yakubovsky2017optical}.

\section{Results}

\subsection{Switching energy}

\begin{figure*}[t]
    \centering
    \includegraphics[width=1\textwidth]{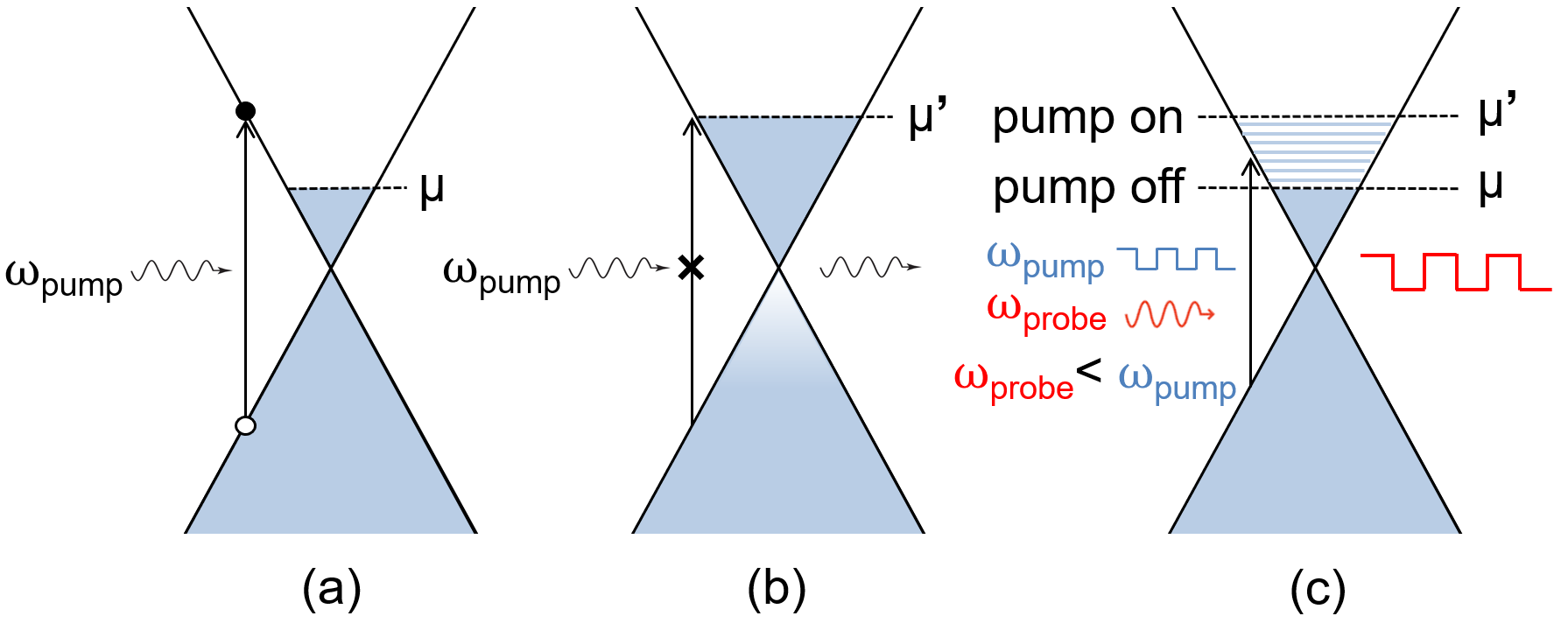}
    \caption{(a) Interband absorption of a pump photon with energy $\hbar \omega_{\text{pump}}$. (b) Pump photon transmitted after applying a sufficiently high pump intensity. (c) Transmission of the probe signal is determined by the pump signal amplitude. Black and white circles represent electrons and holes, respectively. Filled energy states are represented by darker shades. Reprinted with permission \cite{alaloul2021low}, https://pubs.acs.org/doi/10.1021/acsomega.0c06108. Further permission related to the material excerpted should be directed to the American Chemical Society.}
    \label{fig:transitions}
\end{figure*}

The absorption of graphene is tuned by Pauli-blocking \cite{alaloul2021low}, where photogenerated electrons fill the conduction band states of graphene following a sufficiently intense pump excitation, and by that, they block the interband transition of other electrons (see Fig. \ref{fig:transitions}). At near-IR wavelengths, interband absorption is the dominant absorption mechanism in graphene \cite{liu2018graphene}. The interband absorption of graphene saturates when the chemical potential ($\mu$) reaches a value of $\mu^\prime \approx \hbar\omega_{\text{pump}}/2$, as inferred from Fig. \ref{fig:transitions}. Consequently, incoming pump photons cannot induce interband absorption in graphene because $\hbar\omega_{\text{pump}} < 2|\mu^\prime|$ \cite{liu2018graphene}, and are therefore transmitted. Likewise, a probe photon that has an energy $\hbar\omega_{\text{probe}} < \hbar\omega_{\text{pump}}$ is also transmitted because $\hbar\omega_{\text{probe}} < 2|\mu^\prime|$. Making use of the Pauli-blocking physics in graphene enables all-optical switching: a probe signal is transmitted when the pump signal is ON, or absorbed when the pump signal is OFF. The interband absorption is saturated when the carrier density ($\Delta n$) of graphene increases by \cite{alaloul2021low}:

\begin{equation}
    \Delta n = \dfrac{1}{\pi} \left(\dfrac{\Delta \mu}{\hbar v_{F}} \right)^2 \: , \:\:\: \Delta \mu = \mu^\prime - \mu \, ,
\end{equation}

\noindent with $v_F$ being the Fermi velocity. Considering a graphene sheet with an area of $A=W_G L_{\text{sw}}$, the number of electrons that would be needed to reach $\mu^\prime$ is $m = \Delta n W_G L_{\text{sw}}$. The saturation threshold or the switching energy ($U_{\text{sw}}$) can be expressed as \cite{ono2020ultrafast, alaloul2022high}:

\begin{equation}
U_{\text{sw}} = \sum_{m} \hbar\omega_{m} \, ,
\end{equation}

In practice, there are other loss mechanisms that influence the saturation threshold of the device. Hence, the effective switching energy ($U_{\text{eff}}$) is calculated by taking these loss mechanisms into account \cite{alaloul2021low, alaloul2022chip}:

\begin{equation} \label{ueff_energy}
    U_{\text{eff}} = \dfrac{U_{\text{sw}} * (1 + \Gamma + A_{\text{WG}})}{A_{\text{G}}*{(1-A_{\text{ns}})}}
\end{equation}

\begin{figure}
  \centering
  \includegraphics[width=0.6\linewidth]{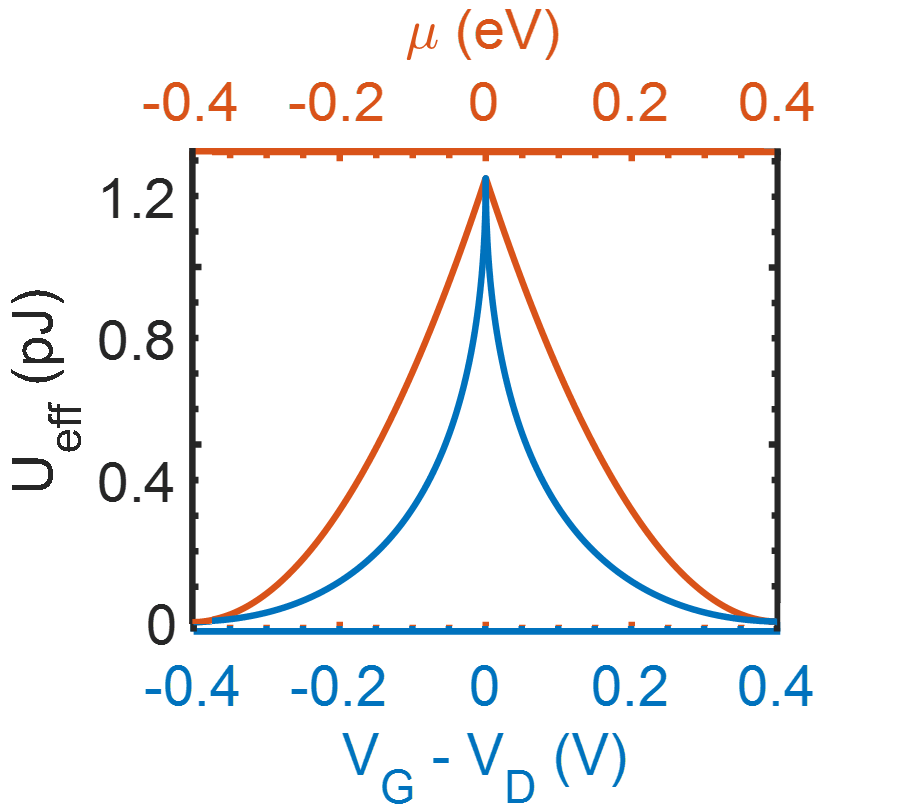}
\caption{Effective switching energy ($U_{\text{eff}}$) as a function of chemical potential ($\text{\textmu}$) and the applied DC voltage ($V_G - V_D$). $\lambda_{\text{pump}}$ = 1550$\,$nm.}
\label{fig:mu}
\end{figure}
\begin{figure} 
    \centering
        \subfloat[\label{fig:circuit1}]{%
       \includegraphics[width=0.5\linewidth]{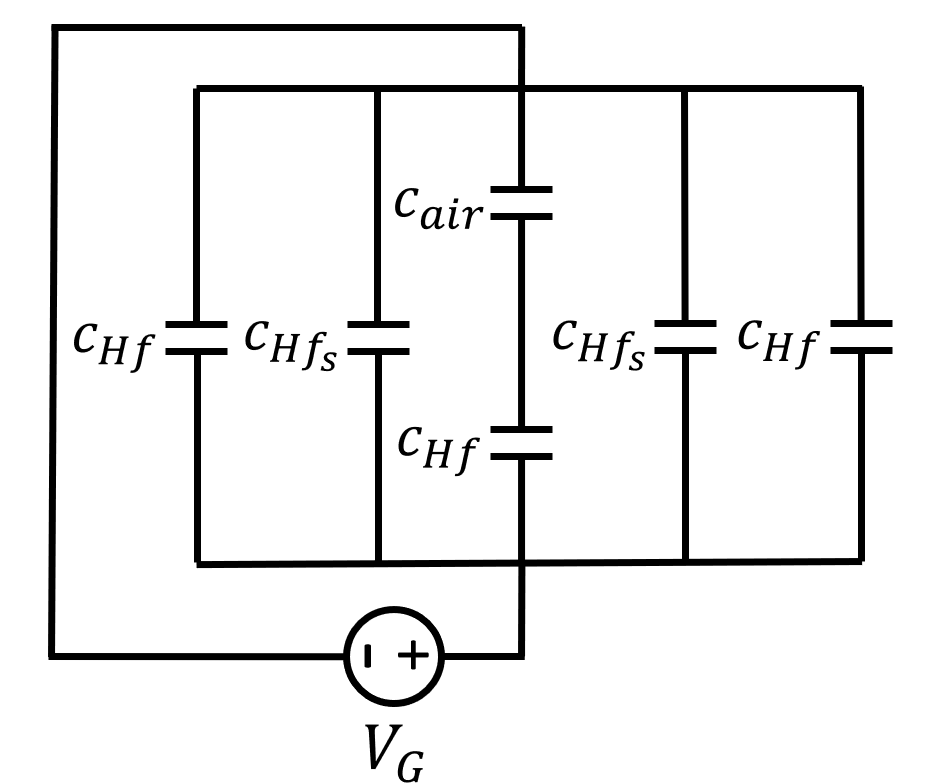}}
       \hfill
       \subfloat[\label{fig:voltage}]{%
        \includegraphics[width=0.5\linewidth]{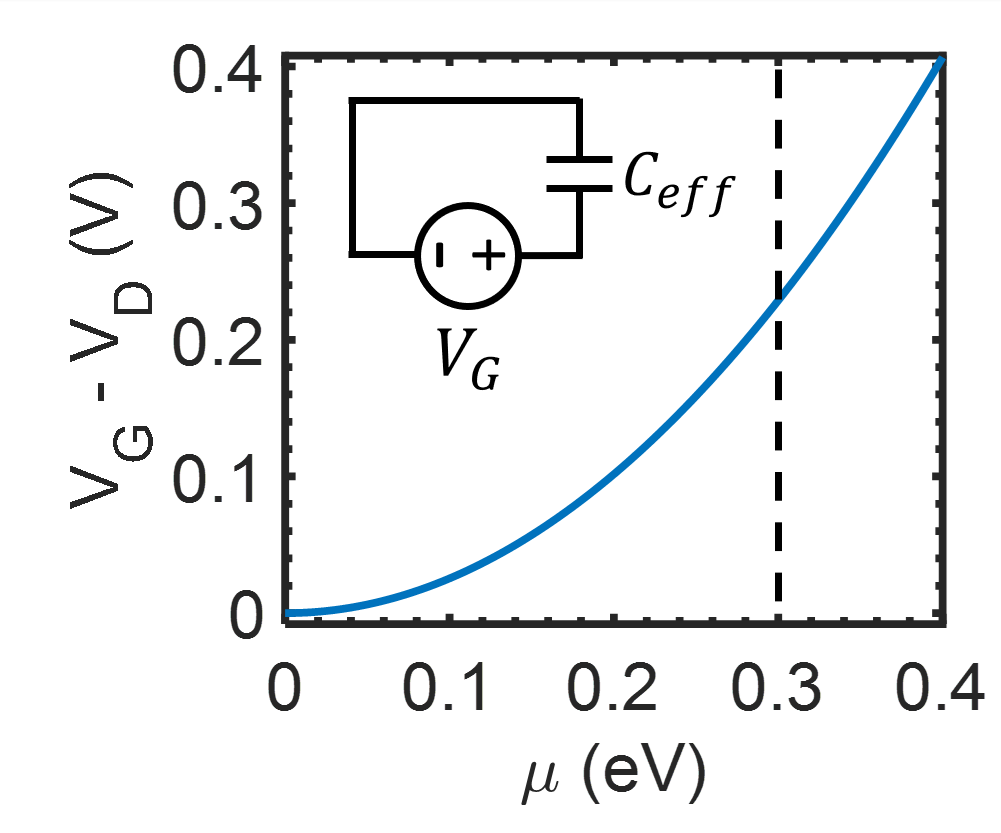}}
  \caption{(a) Equivalent capacitance of the device. $c_{\text{Hf}}$, $c_{\text{Hf}}$, and $c_{\text{air}}$ are the capacitances per unit area of hafnia, side hafnia and air, respectively. (b) Applied DC voltage ($V_G - V_D$) as a function of the chemical potential ($\text{\textmu}$). The inset shows the equivalent effective capacitance.}
  \label{figNew} 
\end{figure}


\noindent with $\Gamma = 1-\kappa$ being the coupling loss, $A_{G} = 1 - 10^{-(\alpha/10)*L}$ is the percentage of light absorbed by graphene, which consists of saturable and non-saturable parts, i.e. $A_{Gr} = A_{Gr}(A_{s} + A_{ns}) = A_{Gr}A_{s} + A_{Gr}A_{ns}$. $A_{\text{ns}}$ is the non-saturable absorption percentage of $A_G$ \cite{bao2009atomic, bao2011monolayer, zhang2015dependence}, which can be taken as 5\% for monolayer graphene \cite{alaloul2021low, bao2011monolayer}. $A_{\text{WG}} = 1 - 10^{-(\alpha_{\text{WG}}/10)*L}$ is the fraction of light that is lost due to the waveguide losses that are not related to graphene, with $\alpha_{\text{WG}} = 7\,$dB/cm taken from the values reported in \cite{serna2014potential}, where a silicon slot waveguide with an 80$\,$nm wide slot is reported. The calculated $U_{\text{eff}}$ is shown in figure \ref{fig:mu} as a function of $\mu$ for $\lambda_{\text{pump}} = 1550\,$nm. As explained in \cite{alaloul2021low}, at low chemical potentials, more electrons would be needed to fill the conduction band states up to $\mu^\prime$, which results in higher $U_{\text{eff}}$. Tuning the chemical potential of graphene is possible by electrostatic doping, which is achieved in our configuration by applying the gate voltage ($V_G$) as shown in figure \ref{1a}. Then, the relation between $V_G$ and $\text{\textmu}$ is given by \cite{opex}:

\begin{equation} \label{eq:relation}
    V_G = \dfrac{e  \mu^2}{\pi C_{\text{eff}} \hbar^2 v_{F}^2} \, ,
\end{equation}

\noindent with $C_{\text{eff}}$ being the effective capacitance per unit area. Figure \ref{fig:circuit1} presents the equivalent capacitance of the device, and the equivalent effective capacitance is shown in the inset of figure \ref{fig:voltage}. In our configuration, $C_{\text{eff}} = 2*c_{\text{Hf}} + 2*c_{\text{Hf$_{s}$}} + (1/c_{\text{air}} + 1/c_{\text{Hf}})^{-1}$, where $c_{\text{Hf}} = \epsilon_0 \epsilon_{\text{Hf}} / d_{\text{Hf}}$ is the hafnia capacitance, $c_{\text{Hf$_{s}$}} = \epsilon_0 \epsilon_{\text{Hf}} / d_{\text{Hf$_s$}}$ is the the capacitance of hafnia on the slot sides, and $c_{\text{air}} = \epsilon_0 \epsilon_{\text{air}} / d_{\text{air}}$ is the air capacitance. From figure \ref{1a}, $d_{\text{Hf}} = 10\,$nm, $d_{\text{Hf$_s$}} = 250\,$nm, and $d_{\text{air}} = 240\,$nm. $\epsilon_{\text{air}} = 1$ and $\epsilon_{\text{Hf}} = 25$ are the dielectric constants of air and hafnia \cite{robertson2004high}, respectively. In figure \ref{fig:voltage}, it is observed that large variations in $\text{\textmu}$ are obtained by applying small bias voltages of $<0.5\,$V. This is because of the high dielectric constant ($\epsilon_{\text{Hf}} = 25$) and small thickness ($d_{\text{Hf}} = 10\,$nm) of the hafnia layer, which results in a high total effective capacitance, leading to a low $V_G$. Using equation \ref{eq:relation}, $U_{\text{eff}}$ is plotted in figure \ref{fig:mu} as a function of $V_{G} - V_{D}$ for $\lambda_{\text{pump}} = 1550\,$nm, with $V_{D}$ being the gate voltage for the chemical potential to be tuned at the Dirac point \cite{liu2018graphene}. $U_{\text{eff}}$ is reduced as $V_G - V_D$ increases, which confirms that the switching energy of the all-optical switch is tuned by electrostatic doping of graphene. As explained in \cite{opex}, this technique is not omnipotent because the non-zero scattering rate of graphene and the ambient operating temperature fundamentally limit the graphene absorption, and consequently the switching efficiency, at high bias voltages (see Supplement 1, Section 3). To operate the device without compromising its switching efficiency at $\lambda_{\text{pump}} = 1550\,$nm and at an operating temperature of 300$\,$K and a 100$\,$fs scattering time, the maximum $V_G - V_D$ can correspond to a $\text{\textmu}$ value of $\sim 0.3\,$eV or less. From figure \ref{fig:voltage}, it can be seen that for $\text{\textmu} = 0.3$, $V_G - V_D = 0.23$. Thus, the minimum attainable $U_{\text{eff}}$ is $\sim 79\,$fJ at $V_G - V_D \approx 0.23\,$V, based on the data presented in figure \ref{fig:mu}. Therefore, ultra-efficient all-optical switching is achieved.

\subsection{Switching efficiency}

The switching efficiency of the device is characterized by the extinction ratio, $ ER = 10\,\text{log}_{10} (T_{\text{on}} / T_{\text{off}} )$, and the losses are quantified by the insertion loss, $IL = 10\,\text{log}_{10} (1 / T_{\text{on}})$ \cite{alaloul2021low}. $T_{\text{off}}$ and $T_{\text{on}}$ represent the transmitted power of the probe signal when the pump signal is turned off and on, respectively. The absorption of graphene is maximized when the pump signal is turned off, whereas the maximum transmission of the probe signal ($T_{\text{max}}$) is attained when a pump signal with an energy $U>U_{\text{eff}}$ is applied. $T_{\text{max}}$ and $T_{\text{off}}$ can be expressed as \cite{alaloul2021low}:

\begin{equation} \label{Tmax}
    T_{\text{max}} = [1 - (\Gamma + A_{\text{WG}} + A_{\text{Gr}}A_{\text{ns}})]* (1-\Gamma)
\end{equation}

\begin{equation} \label{Toff}
    T_{\text{off}} = [1 - (\Gamma + A_{\text{WG}} + A_{\text{Gr}})] * (1-\Gamma)
\end{equation}

\begin{figure} 
    \centering
  \subfloat[\label{ERIL}]{%
       \includegraphics[width=0.5\linewidth]{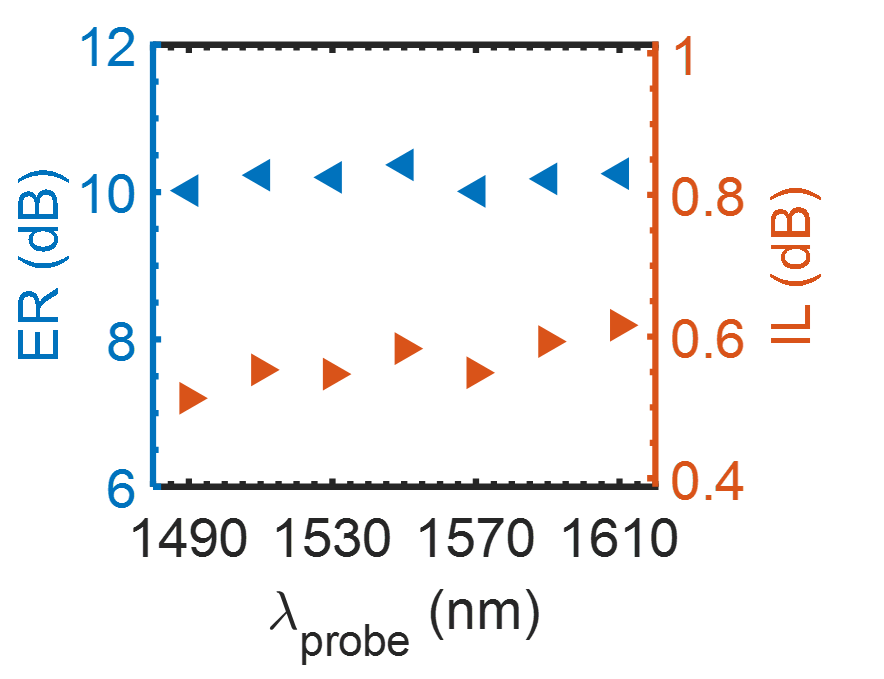}}
    \hfill
  \subfloat[\label{ueff_lambda}]{%
        \includegraphics[width=0.5\linewidth]{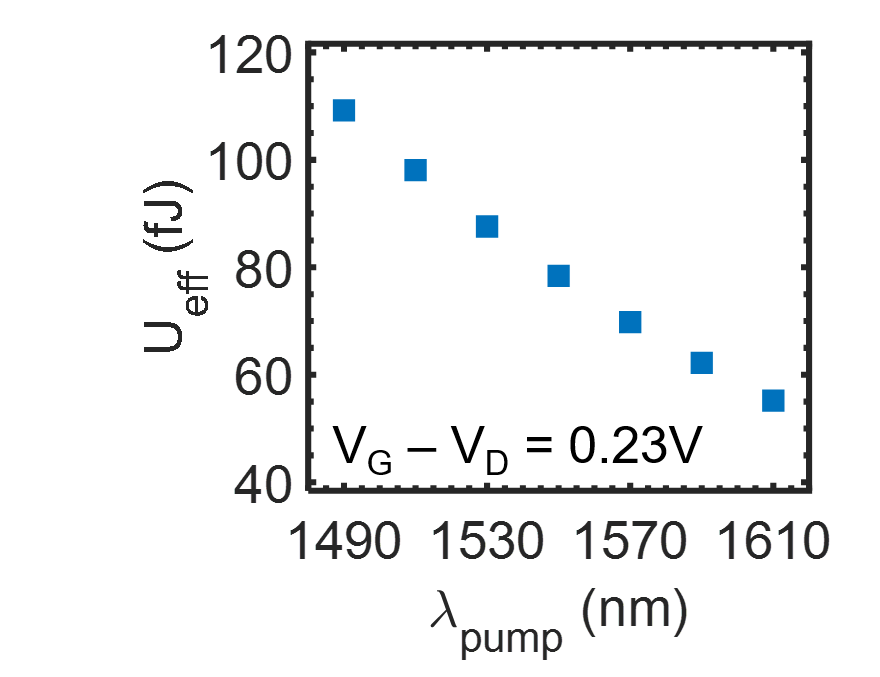}}
  \caption{(a) Maximum extinction ratio ($ER$) and insertion loss ($IL$) as a function of the probe signal wavelength ($\lambda_\text{probe}$). (b) Effective switching energy ($U_{\text{eff}}$) as a function of the pump signal wavelength ($\lambda_\text{pump}$) at $V_G - V_D = 0.23\,$V.}
  \label{ER} 
\end{figure}

Using this device, a high $ER$ of 10.3$\,$dB and an ultra-low $IL$ of 0.58$\,$dB are obtained at $\lambda_{\text{probe}} = 1550\,$nm. The broadband response of the device is quantified by calculating the $ER$, $IL$, and $U_{\text{eff}}$ at other wavelengths, using the computed $\alpha$ and $\kappa$ values from figure \ref{wavelength_sweep}. A similar waveguide loss of $\alpha_{\text{WG}} = 7\,$dB/cm is also used for calculations at other wavelengths. Figure \ref{ERIL} shows the calculated $ER$ and $IL$ as a function of the probe signal wavelength ($\lambda_{\text{probe}}$). It is observed that the $ER$ slightly varies at other wavelengths because of the differences in $\alpha$, as shown in figure \ref{wavelength_sweep}, and as a result, the $ER$ varies  according to Eqs. \ref{Tmax} \& \ref{Toff}. Similarly, the $IL$ slightly varies at other wavelengths because of the variations in the coupling loss, based on the computed $\kappa$ values in figure \ref{wavelength_sweep}. In the studied wavelength band, $A_G \geq 95\,$\% for chemical potentials up to $\sim 0.3\,$eV (see Supplement 1, Section 3). Hence, a similar voltage ($V_G - V_D = 0.23\,$V) is applied at the other wavelengths. Figure \ref{ueff_lambda} presents the computed $U_{\text{eff}}$ as a function of the pump signal wavelength ($\lambda_{\text{pump}}$) at $V_G - V_D = 0.23\,$V. $U_{\text{eff}}$ increases at shorter wavelengths, which agrees with experimental observations \cite{zhang2015dependence}. The device efficiently operates at telecom C-band wavelengths and beyond, and that is desirable for broadband optical communication networks.

\subsection{Switching performance}

The switching performance is quantified by the rise and fall times of the device, which are related to the electron heating and cooling mechanisms in graphene \cite{alaloul2021low, opex}. Because of the conical dispersion of graphene, the density of states fades out at the Dirac point. Therefore, near the Dirac point, electrons have a relatively low heat capacity, and when photoexcited, they instantly scatter with other electrons, thereby creating a momentary sea of hot electrons in $<150\, $fs \cite{tielrooij2015generation, chen2019highly, alaloul2021plasmon}. The generated sea of hot electrons later cools down in a few picoseconds by phonon- and disorder-assisted scattering \cite{song2012disorder, ma2014competing, chen2008charged,graham2013photocurrent}. The device switching time is quantified by first considering the electrical conductivity of graphene \cite{lin2019asymmetric, song2011hot, shiue2015high, alaloul2021plasmonic}:

\begin{equation} \label{eq:conductivity}
    \sigma = \sigma_{0}(1+\dfrac{\text{\textmu}^{2}}{\Delta^{2}}) \; , \; \;\sigma_{0} = 5(\dfrac{e^{2}}{h}) \, ,
\end{equation}

\noindent where $h$ is Planck's constant. $\Delta = 100\,$meV is the minimum conductivity plateau, and $\sigma_{0}$ is the minimum conductivity of graphene; the values of $\Delta$ and $\sigma_{0}$ are taken from \cite{lin2019asymmetric}. The mobility of graphene ($\eta$) can be calculated from $\sigma$ \cite{yan2011correlated}:

\begin{equation} \label{eq:drude}
    \eta = \dfrac{\sigma}{en_0} \, ,
\end{equation}

Using the Boltzmann transport theory, the scattering time is given by  \cite{zhu2009carrier}:

\begin{equation} \label{eq:scat}
    \tau_{\text{scat}} = \dfrac{\mu\eta}{ev_{F}^2} \, ,
\end{equation}

Using the previously calculated quantities, the electron cooling rate ($\gamma_{\text{cool}}$) is calculated using these equations \cite{alaloul2021plasmon, ma2014competing, lin2019asymmetric}: 

\numparts
\begin{equation} \label{eq:4}
    \gamma_{\text{cool}} = \tau_{\text{cool}}^{-1} = b \: (T +\dfrac{T_{*}^{2}}{T}) \, ,
\end{equation}

\begin{equation} \label{eq:5}
     b = 2.2 \: \dfrac{g^{2} \varrho k_{B}}{\hbar k_{F}\ell} \; , \; \; \; T_{*} = T_{BG} \sqrt{0.43k_{F}\ell}  \, ,
\end{equation}

\begin{align}\label{eq:6}
\begin{split}
    g = \dfrac{D}{\sqrt{2 \rho s^{2}}} \;,  \; \; \; \varrho = \dfrac{2\mu}{\pi \hbar^{2} v_{F}^2} \; , \; \; \; k_F = \dfrac{\mu}{\hbar v_F} \;,  \; \; \; k_{F}\ell = \dfrac{\pi \hbar \sigma}{e^{2}} \; , \; \; \; T_{BG} = \dfrac{s\hbar k_{F}}{k_{B}} 
\end{split}
\end{align}

\endnumparts

\noindent where $T = 300\,$K is the operating temperature, $\varrho$ is the density of states, $k_{F}$ is the Fermi wave vector, $k_{F}\ell$ is the mean free path, $g$ is the electron-phonon coupling constant, $T_{BG}$ is the Bloch-Grüneisen temperature, $D=\,$20$\,$eV is the deformation potential constant \cite{song2012disorder}, $\rho=7.6\times10^{-7}\,\text{Kg}/\text{m}^{2}$ is the mass density of graphene, and $s = 2\times10^{4} \, \text{m}/\text{s}$ is the speed of longitudinal acoustic phonons in graphene \cite{liu2018graphene}. The electron cooling time ($\tau_{\text{cool}}$) is then taken as the inverse of $\gamma_{\text{cool}}$. 

A sea of hot electrons is created within a timescale of $\sim \tau_{\text{scat}}$ and fills up the conduction band states, inducing Pauli-blocking. Later, photoexcited electron cool down within a timescale of $\sim \tau_{\text{cool}}$ \cite{opex}, enabling the interband absorption of incoming photons. Therefore, $\tau_{\text{scat}}$ and $\tau_{\text{cool}}$ are taken as the rise and fall times of the switch, respectively. The calculated $\tau_{\text{scat}}$ and $\tau_{\text{cool}}$ are $<150\,$fs and $<5\,$ps, respectively, for $0.1 \leq \mu \leq 0.3\,$eV. Therefore, the exceptionally fast electron heating and cooling dynamics in graphene enable ultra-high-speed switching. 

\section{Discussion}

Table \ref{tab10} summarizes the reported performance metrics of demonstrated on-chip all-optical graphene switching devices. The device presented in \cite{ono2020ultrafast}, is based on graphene-loaded plasmonic slot waveguides. It achieves the highest modulation efficiency (0.875$\,$dB/\textmu m), the lowest switching energy ($35\,$fJ) and the fastest cooling dynamics (260$\,$fs), but is limited by an excessive $IL$. The non-plasmonic devices that were reported in \cite{sun2018all, wang2020cmos, qiu2021high}, have a much lower $IL$ and can achieve high modulation efficiencies at large $L$, but their switching energies may exceed $1\,$pJ/bit at these lengths. Though this device is longer than other devices, it achieves a relatively high modulation efficiency (0.255$\,$dB/$\text{\textmu}$m), energy-efficient switching ($\sim 79\,$fJ) and ultrafast recovery ($<5\,$ps) at an almost negligible $IL$ of 0.58$\,$dB. 

\begin{table}
\label{tab10}
 \caption{\bf Performance metrics of on-chip all-optical graphene switches/modulators} 

\begin{indented}
\lineup
\item[]\begin{tabular}{@{}*{8}{l}}
\br                              
Ref. & $IL$ & $ER$ & $L$ & $ER$/$\mu$m & $IL$/$\mu$m  & $\tau_{\text{cool}}$ & $U_{\text{eff}}$ \\ 
\mr
\cite{ono2020ultrafast} & 19$\,$dB & 3.5$\,$dB & 4$\,\text{\textmu}$m & 0.875$\,$dB/$\text{\textmu}$m & 4.75$\,$dB/$\text{\textmu}$m  & 260$\,$fs & 35$\,$fJ \cr

\cite{sun2019all} & n/a & 2.1$\,$dB & 10$\,\text{\textmu}$m & 0.21$\,$dB/$\text{\textmu}$m & n/a & n/a & n/a \textsuperscript{\emph{a}}\cr

\cite{sun2018all} & negligible & 2.75$\,$dB & 100$\,\text{\textmu}$m & 0.0275$\,$dB/$\text{\textmu}$m & negligible & n/a  & n/a \textsuperscript{\emph{b}}\cr 

\cite{wang2020cmos} & negligible & 1.1$\,$dB \textsuperscript{\emph{c}} & 30$\,\text{\textmu}$m & 0.0367$\,$dB/$\text{\textmu}$m & negligible & 1.2$\,$ps \textsuperscript{\emph{d}} & 1.38$\,$pJ \textsuperscript{\emph{e}} \cr 

\cite{qiu2021high} & $\sim 1\,$dB \textsuperscript{\emph{f}} & 11$\,$dB & 288$\,\text{\textmu}$m \textsuperscript{\emph{g}} & 0.038$\,$dB/$\text{\textmu}$m & 0.0035$\,$dB/$\text{\textmu}$m & 1.29$\,$\textmu s & n/a \textsuperscript{\emph{h}} \cr 

Here & $0.58\,$dB & 10.3$\,$dB & 40$\,\text{\textmu}$m & 0.2575$\,$dB/$\text{\textmu}$m & $0.0145\,$$\,$dB/$\text{\textmu}$m &  $<5\,$ps &  $\sim79\,$fJ \cr

\br
\end{tabular}
\textsuperscript{\emph{a}} 46$\,$mW is the maximum input light power;
\textsuperscript{\emph{b}} 60$\,$mW is the input light power;
  \textsuperscript{\emph{c}} Modulation depth is 22.7\%;
  \textsuperscript{\emph{d}} Setup-limited by the resolution time of the asynchronous pump–probe system;
  \textsuperscript{\emph{e}} Saturation threshold;
  \textsuperscript{\emph{f}} Waveguide loss before transferring graphene.
  \textsuperscript{\emph{g}} Length of the graphene coating on the waveguide.
  \textsuperscript{\emph{h}} 90--$\,$109.6$\:$mW switching power.
  n/a: not available (not reported).
\end{indented}
\end{table}

\section{Conclusion}

\indent To sum up, we present a novel design of an all-optical graphene switch that is integrated into silicon slot waveguides, which enhance the absorption of graphene by their strong confinement of the guided optical mode. In addition, the device is electrically biased to control its saturable absorption threshold and switching energy, leading to ultra-efficient operation. Simulations were performed to optimize the design and to model the device response. Then, the device switching energy, switching efficiency, broadband response, and switching performance were discussed and compared with recently reported devices. The ultrafast response, high extinction ratio, ultra-low insertion loss and energy-efficient switching of this device are highly promising for all-optical signal processing systems.

\section*{Data availability statement}

The data that support the findings of this study are available upon reasonable request from the authors.

\section*{Acknowledgments}
This work was supported by the Australian Research Council (DP200101353).

\appendix

\section*{Appendix A. Slot width}
\setcounter{section}{1}

\begin{figure} 
  \centering
  \includegraphics[width=0.5\linewidth]{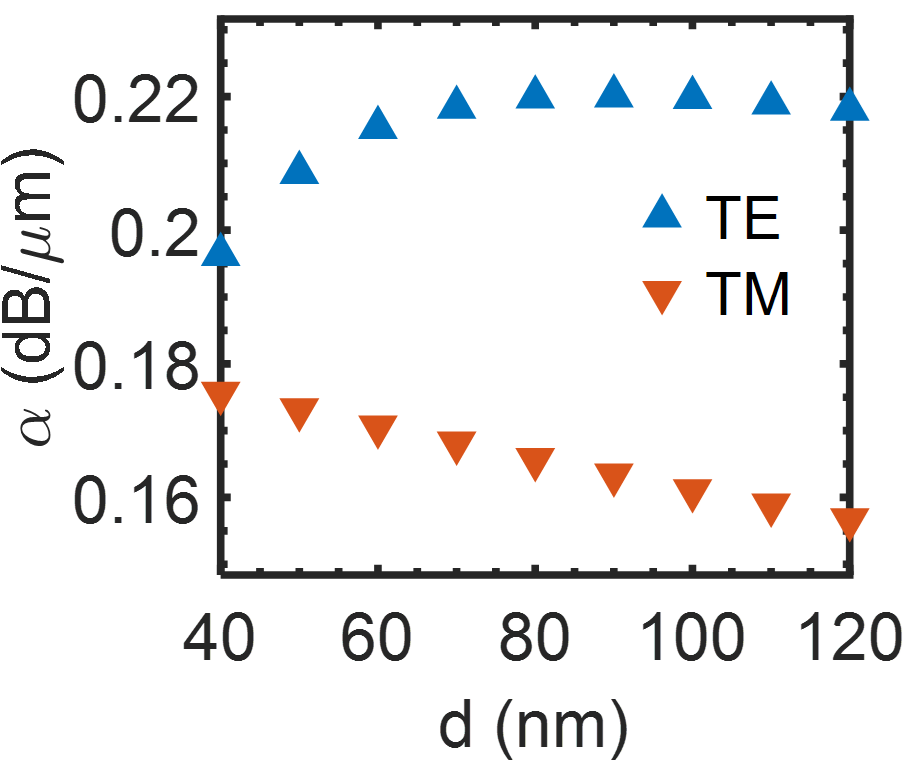}
\caption{Propagation loss ($\alpha$) as a function of the slot width ($d$) for the transverse-electric (TE) and transverse-magnetic (TM) modes. $\lambda=1550\,$nm.}
\label{fig:slot}
\end{figure}

The slot width ($d$) is swept from 40$\,$nm to 120$\,$nm, and the simulated propagation losses ($\alpha$) of the transverse-electric (TE) and transverse-magnetic (TM) modes are recorded (see figure \ref{fig:slot}). It is observed that waveguides with slot widths of 80$\,$nm and $90\,$nm yield the highest absorption. In \cite{serna2014potential}, a Si slot waveguide with a slot width of $80\,$nm has been demonstrated with a low propagation loss of $7\pm2\,$dB/cm. Thus, our device is designed with a similar slot width.

\section*{Appendix B. Effective absorbing width}
\setcounter{section}{2}

Initial simulations were performed with a 3.48$\,\text{\textmu}$m wide graphene sheet as shown in figure \ref{fig:3micron}. The resulting propagation loss ($\alpha$) for the TE-mode is $\sim0.22\,$dB/$\text{\textmu}$m. It might be assumed that the portion of graphene that is on top of the slotted region is the one that solely contributes to the absorption of the waveguide mode, because that is where the waveguide mode is mostly confined. However, when simulating the same structure with an 80$\,$nm wide graphene sheet on top of the slotted region, as shown in figure \ref{fig:80nano}, the resulting $\alpha$ is merely $\sim0.09\,$dB/\textmu m. This indicates that other portions of the graphene sheet significantly contribute to the absorption. Therefore, the width of the graphene sheet is swept to find out the effective absorbing width, which would yield a $\sim 0.22\,$dB/\textmu m propagation loss. It is found out that a 1.64$\,$\textmu m wide graphene sheet yields a $\sim 0.22\,$dB/\textmu m propagation loss (see figure \ref{fig:1400}). Hence, 1.64$\,$\textmu m is taken as the effective absorbing width for calculating the switching energy.

\setcounter{figure}{0}
\begin{figure} 
  \centering
  \includegraphics[width=0.8\linewidth]{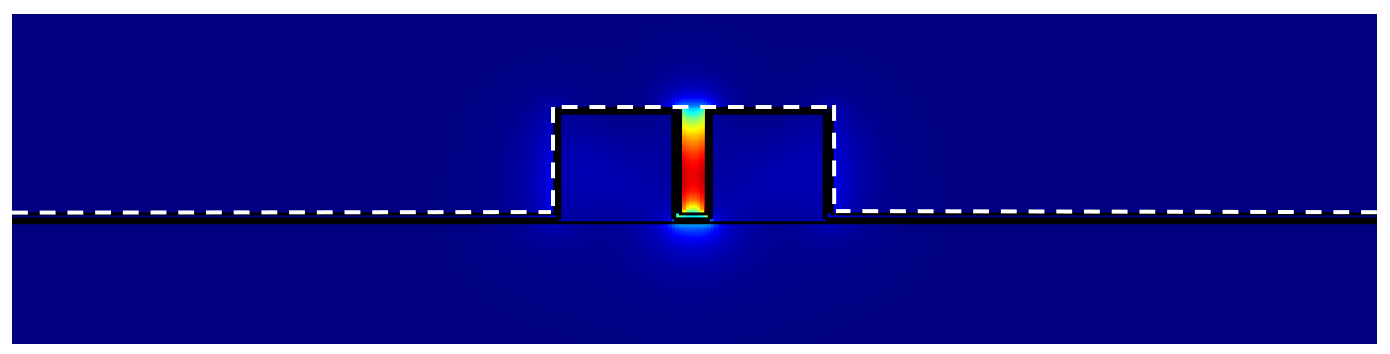}
\caption{Electric field profile of the TE mode for the device with a $3.48\,$\textmu m wide graphene sheet. $\lambda = 1550\,$nm. The white dashed lines represent graphene.}
\label{fig:3micron}
\end{figure}

\begin{figure} 
  \centering
  \includegraphics[width=0.8\linewidth]{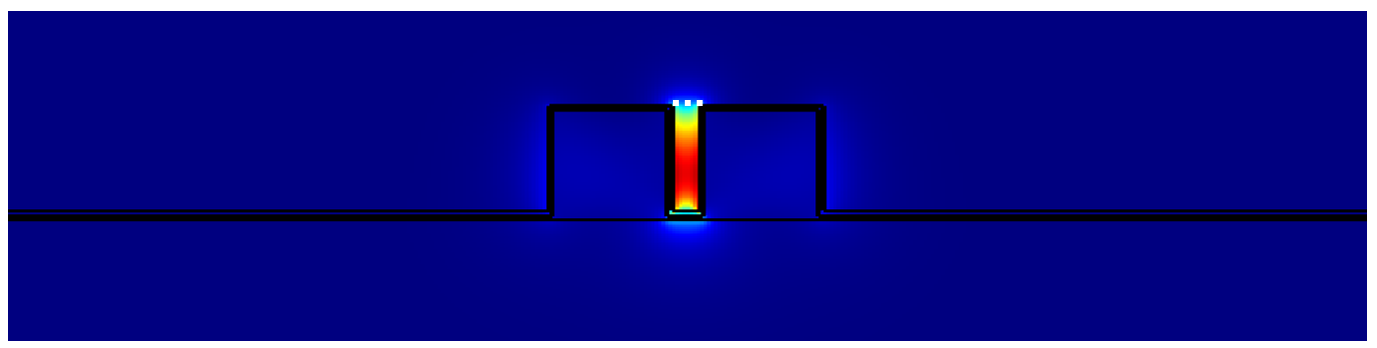}
\caption{Electric field profile of the TE mode for the device with an $80\,$nm wide graphene sheet. $\lambda = 1550\,$nm. The white dashed lines represent graphene.}
\label{fig:80nano}
\end{figure}

\begin{figure} 
  \centering
  \includegraphics[width=0.8\linewidth]{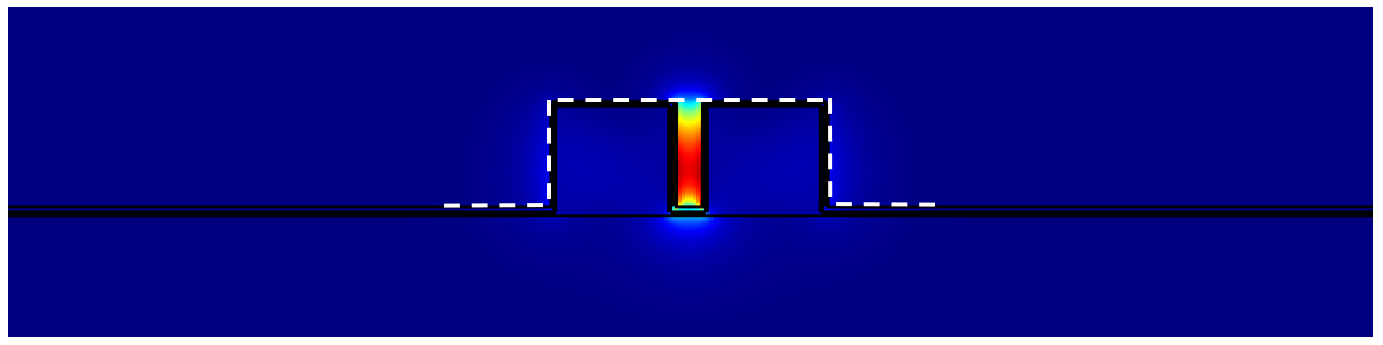}
\caption{Electric field profile of the TE mode for the device with a $1.64\,$\textmu m wide graphene sheet. $\lambda = 1550\,$nm. The white dashed lines represent graphene.}
\label{fig:1400}
\end{figure}

\section*{Appendix C. Contact spacing}
\setcounter{section}{3}

To ensure that the Au/Cr does not induce an ohmic loss, the distance between the metal contacts and Si ralis, denoted by $d_m$, is swept from 0 to 0.8$\,$\textmu m, and the propagation loss ($\alpha$) is recorded. It is noted that $\alpha$ reaches a plateau at large $d_m$, which indicates that the metal contacts are far from the guided mode and cannot induce ohmic losses. A contact spacing of 0.8$\,$nm is chosen because the ohmic losses are eliminated at this distance.

\setcounter{figure}{0}
\begin{figure}
  \centering
  \includegraphics[width=0.5\linewidth]{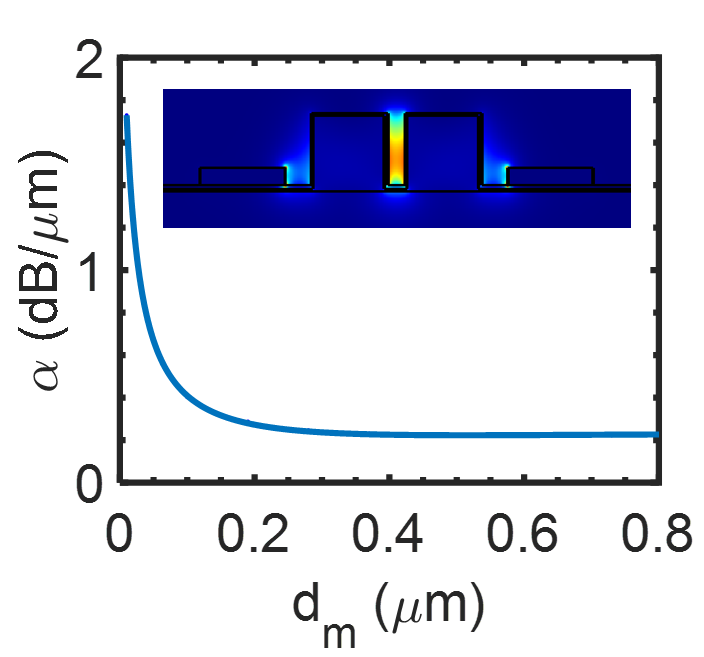}
\caption{Computed propagation loss of the TE-mode as a function of the contacts spacing ($d_m$) from the silicon rails. The inset demonstrates the effect of placing the metal contacts near the guided mode.}
\label{fig:spacing}
\end{figure}

\section*{Appendix D. Graphene absorption}
\setcounter{section}{4}

\setcounter{figure}{0}
\begin{figure}
  \centering
  \includegraphics[width=0.7\linewidth]{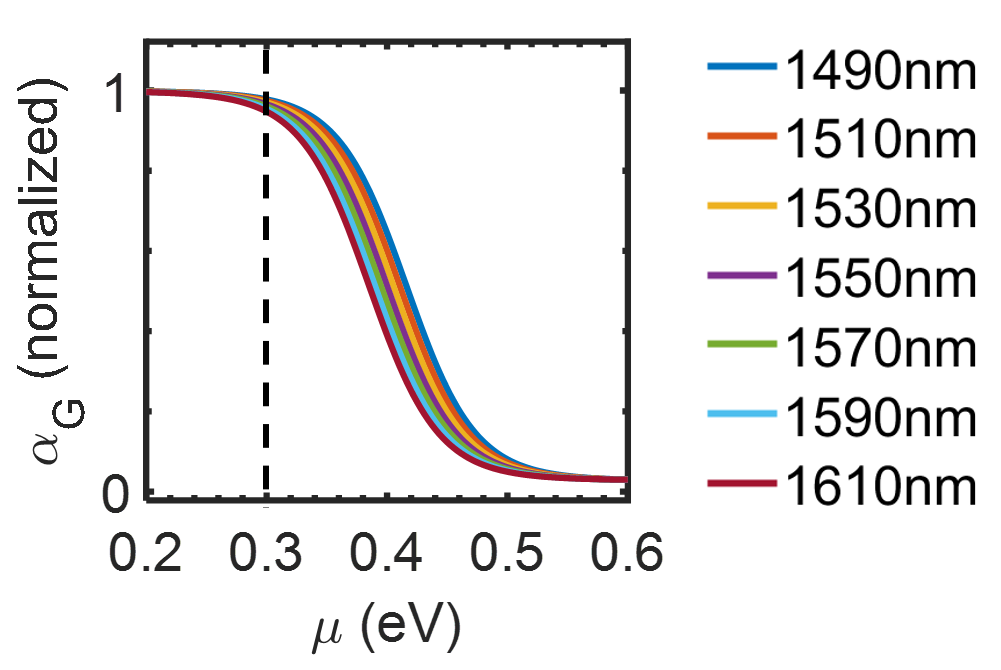}
\caption{Graphene absorption ($\alpha_{G}$) a function of the chemical potential ($\mu$) for multiple wavelengths, at room temperature and for a 100$\,$fs scattering time. }
\label{fig:1610}
\end{figure}

As mentioned in the main text, the absorption of graphene abruptly falls for $\mu > 0.3\,$eV, because of the non-zero scattering rate and the ambient operating temperature (see figure \ref{fig:1610}). As explained in \cite{opex}, $V_G - V_D$ can correspond to a $\text{\textmu}$ value of $\sim 0.3\,$eV or less, to operate the device with a high switching efficiency. The maximum transmittance ($T_{\text{max}}$) in the ON state is $T_{\text{max}} = [1 - (\Gamma + A_{\text{WG}} + A_{\text{Gr}}A_{\text{ns}})]* (1-\Gamma)$, as given in the main text. A reduced graphene absorption ($A_G$) reduces $T_{\text{max}}$, which in turn reduces the maximum extinction ratio ($ER_{\text{max}}$), where $ER_{\text{max}} = 10\,\text{log}_{10} (T_{\text{max}} / T_{\text{off}} )$, as given in the main text. In figure \ref{fig:1610}, it is observed that the absorption falls abruptly for $\mu > 0.3\,$eV, for all wavelengths in the studied band. Figure \ref{fig:1610} is plotted for a temperature of 300$\,$K and 100$\,$fs scattering time, following the steps given in \cite{opex}.

\section*{References}

\bibliographystyle{iop-num}
\bibliography{sample}

\end{document}